\documentclass{article}

{}
{}
{}

\usepackage{amsmath} 
\usepackage{amsthm} 
\usepackage{cite} 
\usepackage{graphicx}
\usepackage[table]{xcolor}

\usepackage{algorithmic} 
\usepackage{authblk}
\usepackage{xspace}
\usepackage[margin=1.2cm]{geometry}
\usepackage{multicol}
\usepackage{blindtext}
\usepackage{tablefootnote}
\usepackage{epstopdf}
\usepackage[hyphens]{url}  
\usepackage[breaklinks=true]{hyperref}
\usepackage{cleveref}

\begin{document}

\title{\textbf{Content ARCs: Decentralized Content Rights in the \\Age of Generative AI}}

\author[1]{Kar Balan}
\author[1]{Andrew Gilbert}
\author[1]{John Collomosse}

\affil[1]{{DECaDE Centre for the Decentralized Digital Economy, University of Surrey, Guildford, UK}\newline }
\date{}

\setcounter{Maxaffil}{0}
\renewcommand\Affilfont{\itshape\small}

\renewcommand{\baselinestretch}{0.95}

\maketitle

\begin{figure*}[t]
    \centering
    \includegraphics[width=\textwidth]{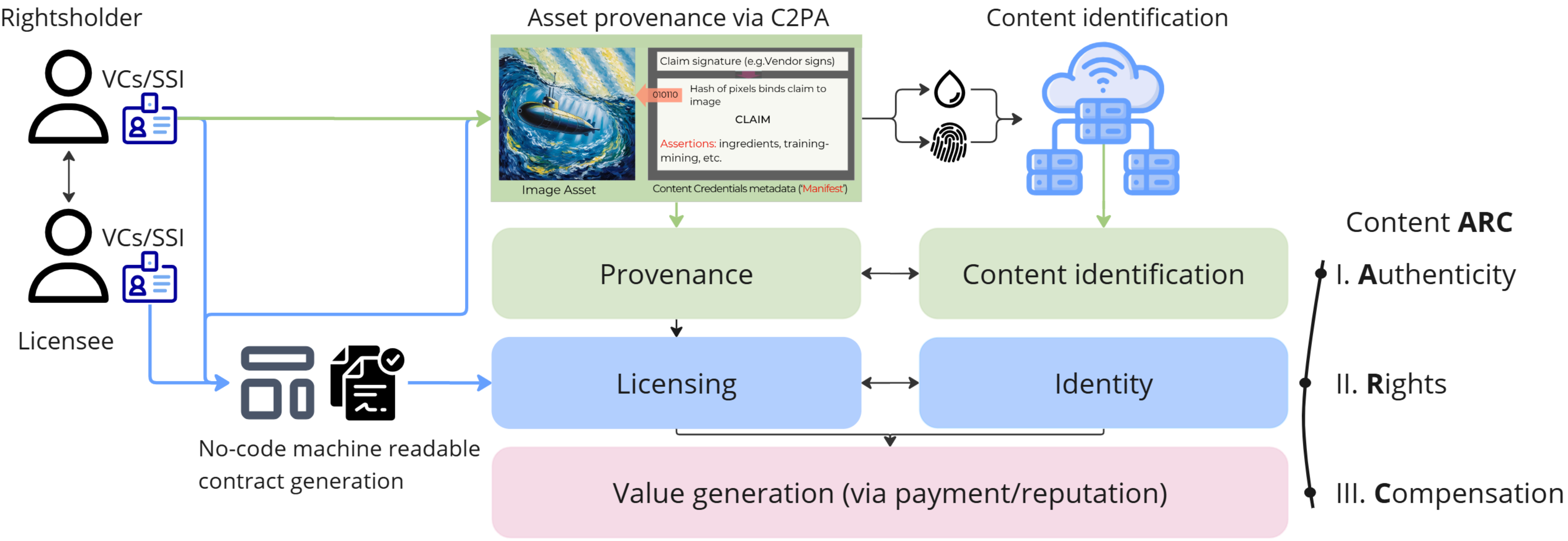} 
    \caption{Content Authenticity, Rights, and Compensation (ARCs) framework, illustrating how content provenance, identification, licensing, and creator identity enable downstream value generation. Green arrows indicate systems feeding into content provenance, while blue arrows feed into licensing.}
    \label{fig:teaser}
\end{figure*}

\begin{abstract}
The rise of Generative AI (GenAI) has sparked significant debate over balancing the interests of creative rightsholders and AI developers. As GenAI models are trained on vast datasets that often include copyrighted material, questions around fair compensation and proper attribution have become increasingly urgent. To address these challenges, this paper proposes a framework called \emph{Content ARCs} (Authenticity, Rights, Compensation). By combining open standards for provenance and dynamic licensing with data attribution, and decentralized technologies, Content ARCs create a mechanism for managing rights and compensating creators for using their work in AI training. We characterize several nascent works in the AI data licensing space within Content ARCs and identify where challenges remain to fully implement the end-to-end framework.
\end{abstract}

\newcommand{\ie}{\textit{i.e.}\xspace}
\newcommand{\eg}{\textit{e.g.}\xspace}

\begin{multicols}{2}

\section{Introduction}

Generative AI (GenAI) is transforming creative workflows through tools for efficiently generating and manipulating text, images, music, and video. However, the vast datasets used to train commercial GenAI models often include copyrighted material \cite{nyt}, causing growing concerns among creative rightsholders about how their work is used and whether they will receive proper recognition and compensation. These concerns have recently surfaced in proposals---and counter-proposals---around changes to copyright legislation in several jurisdictions. The debate has become highly contested; for example, the recent UK copyright consultation \cite{UKAICopyright2025} received over 11,000 submissions following a public awareness campaign by creative practitioners.  Arguments often centre on the contentious issue of opt-in versus opt-out mechanisms; \ie whether permission should always be sought to train commercial AI models on copyrighted data (\ie opt-in) or there should be a default legal right to train unless counter-indicated by a rightsholder (\ie opt-out). Rightsholders often argue that an opt-in model is essential to protect creators’ control over their work and ensure fair compensation \cite{ft2023aicopyright}. While, AI developers often contend that an opt-out approach is necessary given the sheer scale of data (often many billions of assets) over which contractual consent and payment would need to be orchestrated \cite{giustina2024fair}.  Political factors also weigh on the debate, such as creating a competitive AI development environment versus jurisdictions with more permissive legislation or less stringent enforcement \cite{EUAIInvestment2025}. These positions are often deeply polarized and appear difficult to reconcile, highlighting the need for balanced solutions that uphold both creative rights and technological progress.  

In this paper, we contribute a framework to help resolve this deadlock by leveraging a combination of open standards and emerging data-centric technologies such as data attribution and distributed ledger technology (DLT).  We present this as a meta-system design comprising of three core phases (ARC) applied to content: \textbf{(A)uthenticity}, \textbf{(R)ights}, \textbf{(C)ompensation}; hereafter \textbf{Content ARCs}.  Multiple technology choices exist for each phase, and we describe technical options for each.

The motivation of Content ARCs is to provide an automated mechanism for clearing rights to, and compensating rightsholders for, the use of copyrighted data in GenAI training. In this way, concerns around the practicality of opt-in are alleviated through an interoperable machine-actionable protocol leveraging open standards.  While this framework may not always be necessary---AI developers may own, or collectively license \cite{journallicensing}, large datasets to support training---much of the data used for training GenAI models is drawn from the open internet or other fragmented sources where collective licensing is not feasible. In such cases, a scalable mechanism for licensing content from and returning value to individual rightsholders is essential to support large-scale AI development while ensuring fair compensation for creators.  We also discuss legislative support that could benefit the adoption of Content ARCs in the creator economy.

\section{Background}
\label{sec:related} 

Technologies to express creator preferences on AI training or processing are (a) site-based (location-level) or (b) unit-based (asset-level), each approach having strengths and limitations in scope and persistence across the content supply chain.

\noindent {\bf Site-based} methods allow a creator or content host to apply a blanket permission or restriction to assets on all or part of a website.  For decades, the convention (RFC 9309) of parsing a file named `robots.txt' on a server has served as a signal indicating which parts of a website are permitted to be scraped for search indexing. Since different crawlers can be (dis-)allowed per location, and AI training bots self-identify by name, the file can be readily applied for opt-out. Recently, the W3C has proposed a JSON-based variant (TDMRep) \cite{TDMRep2024}, with language aligned to the EU 2019 Copyright Directive \cite{EUCopyright2019}, specifically Article 4, which describes an opt-out mechanism via the Text and Data Mining (TDM) exception.  This use of the term 'Data Mining' here is interpreted as a catch-all for any AI use, reflecting the tension between rapid technology advances and the pace of legislation, which, even in 2019, was too early to consider GenAI explicitly. Ultimately, both approaches allow for efficient expression of opt-in/out in bulk; however, as the signal is not embedded in the assets, it does not persist when content is copied, shared, or aggregated downstream. This limits its use in enforcing opt-in/out within the content supply chain, and there is no mechanism for specifying licensing arrangements for AI re-use.

\noindent {\bf Unit-based} methods allow creators to attach permissions and restrictions directly to individual assets by embedding machine-readable metadata within the file using standards for expressing opt-in/out. Several standards have emerged, ensuring interoperability for this task. The IPTC Photo metadata standard (2024.1) \cite{IPTC2024} includes a flag to indicate 'Data Mining' opt-out, which, similar to TDMRep \cite{IPTC2024}, is a catch-all to indicate opt-out from any form of AI use.  There is no opt-in flag, nor any implication to be drawn as to granting consent in the absence of the opt-out.  Increasingly adopted is the C2PA (Coalition for Content Provenance and Authenticity) \cite{C2PA2024} metadata standard that can also be applied to assets of various modalities (e.g. image, video, audio, text) through embedding of XMP metadata within the file, or a side-car (accompanying linked file) if the format does not support metadata embedding.  C2PA is primarily designed to communicate provenance information \ie an audit history of how an asset is created and supports indication of opt-in/out consent within that audit trail.  Furthermore, opt-in/out may be specified separately for granular AI uses (\eg training or inference) and for generative, non-generative AI, or non-AI analytics (data mining). The metadata is also cryptographically signed to create a non-repudiable, tamper-evident record attached to the asset.

\noindent {\bf Content Identification}. The persistence of unit-based metadata through the content supply chain is a challenge for emerging standards like C2PA that have not reached ubiquity in their adoption.  Many content platforms (particularly social platforms commonly used to redistribute content) strip such metadata from assets.  Therefore, retaining a copy of metadata (including opt-in/out preferences) within a registry is common, coupled with a content identification (Content ID) scheme for looking up that metadata. Examples include invisible watermarking techniques \cite{trustmark,fernandez2024videoseal} that actively inject an identifier into content and content fingerprinting (perceptual hashing) techniques that passively enable identification \cite{Black2021DeepImageComparator, ISCC, nguyen2021oscar}. DECORAIT\cite{balan2023b} combines fingerprinting with a decentralized metadata registry held on a DLT to look up the AI opt-in/out status of images, expressed in C2PA metadata, while IPTC-PLUS \cite{plus} used fingerprinting to look up the opt-out status of images expressed in IPTC metadata.

\noindent{\bf Digital Rights Management (DRM)} refers to technical mechanisms which enforce access control over digital content. Unlike opt-in/out signals, which are primarily declarative forms of usage right, DRM actively prevents unauthorized use or copying of content through encryption and other safeguards. In Content ARCs, we suggest licenses for content use and value creation and assert that such licenses should be offered legal support. We do not suggest using them as an access control mechanism, as with DRM.

\section{Findings}

We propose a framework called \emph{Content ARCs} (Authenticity, Rights, Compensation) to address the growing challenges around fair compensation and proper attribution for AI-generated content. Content ARCs aim to provide a scalable, interoperable mechanism for clearing rights and compensating rightsholders for using copyrighted data in GenAI training. As illustrated in \cref{fig:teaser}, the framework is structured around three interconnected phases---Authenticity, Rights, and Compensation---all supported by open standards and decentralized technologies. This design allows for automated rights management and value creation at scale, even when content is distributed or aggregated across multiple platforms.  We outline the key function of each phase and discuss technology choices available for their instantiation.

\noindent \textbf{(A)uthenticity} phase provides a verifiable way to establish the identity and verify the authenticity of a digital asset.

\noindent \textbf{(R)ights} phase provides a way for a rightsholder to assert ownership of a digital asset and to create digital contracts to license that asset for use by others.  Underpinned by digital identity and representations of rights.

\noindent \textbf{(C)ompensation} phase enables rightsholders to extract value when a licensee exercises a granted license.  This includes potentially determining the level of payment via automated means \eg smart contracts and data attribution technologies.

\subsection{Authenticity}
\label{sec:auth}
To license and enable value creation from media, there must first be a basis for identifying assets and establishing their authenticity \cite{Collomosse2024}. 

The Coalition for Content Provenance and Authenticity (C2PA standard) is a cross-industry group that has developed an open standard for encoding provenance metadata directly into media files to help determine their authenticity. The C2PA standard \cite{C2PA2024} defines a data structure known as a manifest, which records key facts---called assertions---about the origin and edit history of an asset. These assertions can include details such as the tools used and any source assets (or ingredients) incorporated into the work. Ingredients themselves can carry C2PA manifests, creating a structured provenance graph. Assertions are organized into a claim, which is cryptographically signed using Public Key Infrastructure (PKI) to ensure authenticity and prevent tampering. The only mandatory assertion is a cryptographic hash of the asset which securely links the manifest to its content.  By examining the provenance of an asset, an end-user may draw a conclusion regarding its authenticity.  Recently, C2PA version 2 has been fast-tracked as an international standard (ISO/DIS 22144), and JPEG Trust also adopted version 1 as part of its international standard (ISO/IEC 21617-1:2025).  JPEG Trust \cite{jpegtrust2025} is technically interoperable with C2PA and re-articulates several C2PA constructs \eg a `Manifest' is referred to as `Trust Manifest'.  It introduces adjunct constructs such as the `Trust Profile', a JSON-based schema for checking if assertions in an asset match some validation criteria. It may be used to help automate authenticity decisions.  Several other metadata standards (IPTC, EXIF) exist for embedding general information at the time of creation, although these are not explicitly focused on authenticity, nor are they cryptographically signed. C2PA has seen strong adoption in the image modality with several camera manufacturers, digital tools (such as Adobe Photoshop and Microsoft Designer) and most Generative AI models (such as DALL-E and Adobe Firefly), and some early adoption in the video space \cite{NSA2025}.  

Metadata schemes provide a way to embed a unique identifier (GUID) in an asset, for example: a Manifest ID in C2PA; or EIDR, for IDs in audiovisual content.  Critics of metadata schemes,  point to their fragility due to metadata stripping on content platforms (c.f. \cref{sec:related}), motivating the need to ground the identifier in a durable identifier derived from content itself (`content ID'), such as a watermark or fingerprint.  In contrast to the open nature of metadata standards, content ID schemes are often proprietary.  This presents a challenge to ARCs in enabling a low-friction, interoperable solution for robust decentralized rights management. Recently, the International Standard Content Code (ISCC)\cite{ISCC} became an international standard (ISO 24138:2024), describing within a standard method to encode `image codes' using the perceptual hash (pHash \cite{zauner2010implementation}); a frequency based (DCT) algorithm.  This may be modified from its usual 64 bits up to 256 bits by concatenating the result computed over a $2\times 2$ spatial grid \cite{ISCCwebsite}.  The cryptographic `birthday attack' estimates the probability of a collision in $n$ bits as $P(n) \approx 1 - e^{-\frac{n^2}{2^{65}}}$ \ie a 50\% chance at only 5 billion assets for 64 bits, making such a facility important at global scale (\ie over trillions of assets). Other open technologies based on DCT are PhotoDNA \cite{photodna} (144 bits, mainly used for CSAM identification) and PDQ \cite{pdq2019} (256 bits). In general DCT hashes exhibit a high false positive matching rate \cite{nguyen2021oscar}, producing false matches under small manipulations. Contemporary methods for content ID rely upon a trained neural network (AI) to derive a descriptor \eg \cite{liu2016deep,nguyen2021oscar,Black2021DeepImageComparator}. Herein lies a further paradox. Open standards are the only practical solution to adoption and interoperability.  However the pace of standardization is at odds with fast moving  technologies like content ID; and functionality defined by a trained AI model is difficult to standardize. Similar arguments may be made for watermarking, where methods based on frequency domain processing are being superseded by AI approaches \cite{trustmark}. This is leading toward de facto standardization where open sourced content ID models are being commercially adopted (\eg TrustMark \cite{trustmark}), creating silos of interoperable solutions.  Some approaches combine both watermarking and perceptual hashing for additional durability \cite{Collomosse2024}. Whilst the above focused on images, the same holds for other modalities.

Registries play a critical role in this phase by linking a content ID (the key), to  a copy of authenticity metadata \eg C2PA or IPTC (the value).  A centralized global registry (`key-value store') is impractical, although registries do exist in specific verticals (like EIDR for advertising).  A federated registry structure with a unified query point could allow AI developers to identify and verify the authenticity of content before using it for model training.   Distributed Ledger Technology (DLT), colloquially referred to as `blockchain,' provides a way to create a decentralized key-value store without relying upon the trust of a single actor or centralized oversight. The scalability of DLT has improved with more recent proof-of-stake (PoS) networks delivering lower cost and latency.  

\subsection{Rights}

The rights phase defines the legal and contractual terms under which content can be used, and requires a non-repudiable mechanism for a rightsholder to issue licenses to use an asset.

Some metadata standards covered within subsec.~\ref{sec:auth} contain a rudimentary rights declaration to indicate opt-in/out of AI asset use. IPTC has a Data Mining opt-out field. C2PA has a Training and Data Mining assertion that can indicate opt-in/out of different processes covering inference vs.\ training, and generative vs.\ non-generative AI use\footnote{In C2PA v2 the AI opt-in/out assertion was moved from the core specification into a community extension called CAWG.}. Yet none of these standards are aimed at comprehensive rights representation, and a more general rights description language would be required for even more granular permissions (\eg specific people, or sector-specific uses). As such, Content ARCs leverage open rights expression standards to communicate these terms in a machine-readable format. Existing standards include the use of W3C RDF by Creative Commons \cite{CreativeCommons2008}, and the ODRL (Open Digital Rights Language) \cite{odrl2018vocab}. Unlike traditional static licensing models, ODRL-based licenses can be updated dynamically as licensing terms change. This implies the need for a further mechanism, external to the asset itself, to track the issuance and revocation of licenses in a non-repudiable way.

One such mechanism is the use of digitally signed license files. These could be distributed alongside the asset itself or stored externally in a registry. A digitally signed file provides cryptographic proof of authenticity and integrity, ensuring that the terms of the license are tamper-evident and traceable to the issuing rightsholder. A DLT-based licensing registry could be coupled with ODRL to allow rightsholders to update licensing terms dynamically, including revocation or conditional licensing (e.g., time-limited or use-limited licenses). Much as with content ID use cases, a DLT based registry provides a decentralized, transparent, and tamper-evident mechanism for representing license issuance.

Digital identity is a key consideration of ownership and licensing, vouching for a physical person as a digital token.  Many centralized identity systems exist, often grounded in government identity document (or 'know your customer' practices).  Decentralized identity schemes gaining traction are verifiable credentials and decentralized identity.  Yet the systems remain technically complex and not widely adopted by end-users, including in the creative sector.  Since rights systems are often coupled with DLT registries, the public wallet address of the DLT is often used to double as a user identifier.

While provenance has seen increasing standardisation through initiatives like C2PA, no equivalent effort yet exists for machine-readable rights and compensation. We highlight the need for coordinated standardisation across these phases to enable end-to-end interoperability and scalable, decentralised licensing for content.
 
\subsection{Compensation}

The Compensation phase addresses how creators are compensated for the use of their content in AI training and downstream value creation. Compensation models in Content ARCs are based on licensing terms defined in the Rights phase and enforced through automated or manual payment mechanisms. In contrast to historic Digital Rights Management (DRM) solutions that gate access, Content ARCs seek to empower creators through flexible licensing relying upon contractual terms for enforcement.  Banks of standardized, vetted contracts could in the future reduce friction for creative practitioners. For example, a creator could select a pre-configured contract for AI training, specify terms (e.g., duration, permitted uses).  Compensation models are coupled to value in the content supply chain, which remains under-researched in the age of generative AI \cite{DigitalCatapult2024}, but may include:

{\bf Royalty or Event based models}.  Smart contracts automate royalty distribution whenever a licensed work is accessed, replicated, or used in AI pipelines. For instance, NFTs that represent licenses and encode licensing terms in associated SCs can trigger a micropayment to the content owner each time a dataset is downloaded or a derived asset is sold \cite{Collomosse2024}.

{\bf  Attribution-Driven Payouts.} The training datasets responsible for creating AI models can be described via provenance metadata attached to models, raising the opportunity to pay royalties to contributors for use of the model \eg when synthetic assets are generated \cite{Collomosse2024}. However, for Generative AI the scale of such datasets runs into the billions, presenting a challenge to fractional compensation. Data attribution can help identify the subset of training data most influential in a given synthetic asset, to target payout.  Several techniques have measured attribution as visual correlation between output and training data  \cite{Balan2023,wang2023evaluating}. Yet correlation does not equal causation. Causative methods have been proposed, such as invisible watermarking in ProMark\cite{asnani2024promark} to measure which mixture of watermarks exists in the output, as well as Shapley value-based analytics \cite{wang2024economic}, however neither scales to billions of data items. Scaling causative attribution to billion-scale datasets for GenAI models remains an open research challenge in the development of royalty mechanisms for GenAI.

{\bf Non-Financial Incentives.} Beyond direct monetary remuneration, compensation may include access to AI tools, promotional visibility (\eg search ranking reward), future licensing discounts or community incentives.

\begin{table*}[t!]
\centering
\renewcommand{\arraystretch}{1.2}
\resizebox{\textwidth}{!}{%
\scriptsize
\begin{tabular}{|p{1.3cm}|p{2.6cm}|p{3cm}|p{3cm}|p{2.4cm}|p{3cm}|p{2.5cm}|}
\hline
\multicolumn{1}{|c|}{\cellcolor{gray!20} \bf Method} & \multicolumn{2}{c|}{\cellcolor{green!20} \bf I. Authenticity} & \multicolumn{2}{c|}{\cellcolor{blue!20} \bf II. Rights} & \multicolumn{2}{c|}{\cellcolor{pink!20} \bf III. Compensation} \\ 
\hline
\multicolumn{1}{|c|}{} & \cellcolor{green!15} \bf Content ID & \cellcolor{green!15} \bf Verification & \cellcolor{blue!15} \bf Representation & \cellcolor{blue!15} \bf Identity & \cellcolor{pink!15} \bf Attribution & \cellcolor{pink!15} \bf Value Exchange \\ 
\hline
EKILA (ORA) \cite{Balan2023} & \cellcolor{yellow!20} C2PA soft binding (fingerprinting and/or watermarking). & \cellcolor{yellow!20} Cryptographically signed provenance (C2PA).& \cellcolor{gray!20} NFTs for licenses expressed in natural language. &\cellcolor{gray!20} Ethereum wallet address.  & \cellcolor{yellow!20} Proportionate attribution via fingerprint for downstream compensation. & \cellcolor{yellow!20}  Crypto-currency micropayment via SC.\\

\hline

Ocean \newline Protocol & Not implemented at the unit (asset) level.& Not implemented. & \cellcolor{yellow!20} Data NFTs (ownership) + Datatokens (access rights as ERC-20 sub-licenses). & \cellcolor{gray!20} Ethereum \& EVM compatible network wallet address. & Not implemented. & \cellcolor{yellow!20} Datatokens (ERC-20) via SC.  \\

\hline

Story \newline Protocol & Not implemented. Supports watermarked asset specified in metadata. & \cellcolor{gray!20} JSON metadata file and Proof of Creativity (IP provenance graph). & \cellcolor{yellow!20} IP asset as NFT (ownership) + License Tokens as NFTs (licensing agreements). & \cellcolor{gray!20} Story wallet address. & \cellcolor{yellow!20} Derivative works tracking and fractional royalties distribution through License Tokens. & \cellcolor{yellow!20} Royalties distributed via SC in native IP token. \\

\hline

Vana \newline Protocol & Not implemented. & Attestations for data quality, but authenticity is not considered. & Tokens represent fractional ownership and governance of DataDAO. & \cellcolor{gray!20} Vana wallet address. & Not implemented. & Distributed via SCs in native VANA token, but only for top 16 DataDAOs. \\

\hline

SongBits & Not implemented. & Not implemented. & NFTs represent shares of royalty rights. & \cellcolor{gray!20} SUI wallet address, no additional guarantees for artist identities. & Not implemented. & \cellcolor{yellow!20} Distributed via SCs in native SUI token. \\

\hline

JPEG \newline Trust \cite{jpegtrust2025} \newline Draft v2& \cellcolor{yellow!20} C2PA soft binding (fingerprinting and/or watermarking). & 
 \cellcolor{yellow!20} Cryptographically verifiable provenance information through the Trust Profile (JSON-based schema). & \cellcolor{yellow!20} Open digital rights language (ORDL) and Trust Manifest checking. Rights registry. & \cellcolor{yellow!20} Verifiable Credentials / DIDs (CAWG). & Not implemented.  & Not implemented. \\

\hline

Fox \newline Verify & \cellcolor{yellow!20} Cryptographic hashing and fingerprinting. & \cellcolor{yellow!20} Cryptographically signed provenance data (non-standard). &  \cellcolor{yellow!20} Licenses are implemented as logic within SCs. & \cellcolor{gray!20} Custom identity registry SC links cryptographic key pairs to real-world identities.  & \cellcolor{gray!20} Partial implementation via ContentGraph and perceptual hash, but no automated downstream compensation. & \cellcolor{gray!20} License sales via SC in MATIC (Polygon DLT) token, no downstream royalties.\\

\hline

\end{tabular}
}
\caption{Existing creative rights management and monetization systems mapped to the Content ARCs framework across its three core phases: Authenticity, Rights, and Compensation. Smart contract is abbreviated as SC.  Yellow indicates component is present in solution.  Gray indicates partially present. White indicates absent.}
\label{tab:arc_methods}
\end{table*}

\section{Discussion}

We discuss how the landscape of nascent decentralized content licensing implementations map onto the Content ARCs framework and summarize relevant systems in table \ref{tab:arc_methods}.

\textbf{Non-Fungible Tokens (NFTs).} There has been a growing trend in the art world to represent physical art as NFTs, offering a digital tokenized version to facilitate trading and verification of ownership. These NFTs are often accompanied by different types of certificates of authenticity. This smart-contract-based trading approach also enables royalties from secondary sales, providing greater control over intellectual property rights and revenue streams, enhancing transparency and fairness for creators. Verisart (\url{www.verisart.com}) verifies artist identity through 3rd party government-issued ID checks and registers digital certificates of authenticity for both physical and digital art on the Bitcoin blockchain. Arcual (\url{www.arcual.com}) issues digitally signed certificates of authenticity and integrates smart contracts to automate licensing agreements and royalty distribution, ensuring artists receive compensation from secondary sales. Artory (\url{www.artory.com}) partners with experts to verify artwork provenance and register cryptographically signed records, additionally incorporating public auction data. Artclear (\url{www.artclear.com}) fingerprints physical artworks using microscopic imaging which records unique surface characteristics. The resulting digital fingerprints, along with associated metadata, are immutably stored on a blockchain. Comparing new scans with registered fingerprints enables the verification of the artwork's authenticity.  NFT platforms rarely address the topic of copyright or rights assignment. Some platforms offer downstream compensation for resale of assets within the same platform, but downstream compensation for general re-use (including for AI training) is unaddressed.

\textbf{EKILA and the ORA Framework} \cite{Balan2023} is an early technical framework and instantiation of mechanisms that aligns with all  of Content ARCs but does not fully deliver across all components. It enables recognizing and rewarding data contributors to GenAI model training using a combination of the C2PA standard and NFTs stored on the Ethereum DLT to create a technical framework known as ORA (Ownership-Rights-Attribution). ORA instantiated the authenticity phase of Content ARCs by leveraging C2PA to record verifiable provenance of digital assets, which in turn can leverage C2PA soft bindings (watermarking and/or fingerprinting) to identify content. By design, C2PA communicates neither the ownership of an asset, nor (beyond the ability to express opt-in/out permissions for AI) usage rights or licensing. Rights and ownership are therefore represented as NFTs, creating an on-chain, machine-readable licensing system. Ownership is expressed by minting the C2PA-signed asset as a regular NFT. Token-based licenses are issued and managed via smart contracts operated by the rightsholder. However ORA lacks standardized rights representation, specifying only that license detail may be expressed as free-text with the NFT. For compensation, ORA hard codes an API to issue royalty payments for content reuse (in particular, for GenAI) through micropayments into a stored value system enabled via smart contracts. Given a synthetic image created by a GenAI model, EKILA applies data attribution using a correlation (fingerprint similarity method) to identify the subset of training images most responsible for that generated image and issue micropayment royalties to their owners. 

\textbf{Ocean Protocol} (\url{www.oceanprotocol.com}) is a decentralized data exchange platform that leverages DLT to facilitate secure and transparent data sharing and monetization. Ocean Protocol publishes datasets as data NFTs, representing entire datasets rather than individual data points. While the protocol addresses certain aspects of the Content ARCs framework, it does not fully implement all phases. Data authenticity is not implemented—the blockchain records dataset transactions, however provenance of individual data points is not verified prior to being ingested into the system. Data owners hold data NFTs representing copyright or exclusive license for the data asset. Access to datasets is granted through datatokens—fungible ERC-20 tokens that function as sub-licenses, enabling monetization. Smart contracts enforce terms of use, but no attribution mechanisms are implemented, with data owners only receiving compensation when consumers purchase their datatokens for data access.

\textbf{The Story Protocol} (\url{www.story.foundation}) provides a DLT-based solution for managing the lifecycle of intellectual property assets, allowing creators to register, license, and monetize their works through smart contracts. Story provides solutions that intersect with each of the Content ARCs phases, albeit with distinctions and areas for further alignment. Each IP asset is represented as an NFT, establishing proof of ownership and origin since ingestion into the system. The Story Protocol enables derivative works to be linked back to the original, establishing a chain of provenance through the Proof of Creativity mechanism. The protocol supports uploading JSON metadata for an IP asset according to a specific structure, providing provenance details like creation timestamp, author, and media type. It also supports the registration and display of AI pipeline metadata, enabling tracking of relationships such as those between models and datasets (e.g., trained\_on, finetuned\_from). However, authenticity of assets is not verified based on this information prior to being ingested. For rights management, rightsholders define usage rights, royalties, and derivative work restrictions via smart contracts which dsitribute License Tokens and automatically enforce these terms. For the compensation phase, the Royalty Module facilitates smart contract-based royalty distribution, supporting complex royalty flows, such as compensating original creators when derivative works generate revenue.

\textbf{The Vana Protocol} (\url{www.vana.org}) is a DLT-based system for AI training data ownership, governance and monetization via DataDAOs and tokenized incentives. While it aligns with aspects of the Content ARCs framework, limitations exist in data authenticity, individual control, and attribution. A Proof-of-Contribution mechanism assesses data quality and generates off-chain attestations linked on-chain. However, it does not verify that data's provenance or authenticity. Data contributors add validated data to DataDAOs in exchange for tokens, which grant voting power. Data usage rights are determined collectively rather than individually. Contributors cannot define granular licenses, as governance decisions are based on majority rule within the DataDAO. Users earn tokens for contributing data and if their data is used in AI training. However, rewards are only distributed to the top 16 Data Liquidity Pools (DLPs) each epoch, introducing compensation uncertainty and encouraging centralization in larger DLPs.

\textbf{SongBits} (\url{www.songbits.com}) introduced a model for music ownership that directly involves fans in the financial success of artists. While it implements limited rights and compensation mechanisms within the Content ARCs framework, it lacks content authenticity or artist identity verification. Artists sell a share of their song's royalty rights as digital 'bits' (NFTs), representing a fractional share of the song's revenue. For rightsholder compensation, SongBits manages the collection of royalties generated from streaming platforms and distributes these earnings proportionally to shareholders. 

\textbf{JPEG Trust} \cite{jpegtrust2025} is an international standard (ISO/IEC 21617-1) which, in version 1, addresses content provenance and authenticity. Its Media Tokenization group is drafting a version 2 of the standard that develops support for Identity and Rights Declaration on top of C2PA.  JPEG Trust is compatible with with C2PA, leveraging its soft binding (watermarking and fingerprint) capabilities for content-dependent identification and downstream attribution. Integrating the CAWG identity assertions, it uses decentralized identifiers (DIDs) for identity.  Attribution is referred to in the context of attributing rights, rather than as fractional compensation scheme, using machine-readable rights expressed via the ODRL \cite{odrl2018vocab}. These are encoded within the image metadata and could therefore facilitate automated rule-based micro-licensing, though specific compensation mechanisms are not defined. Details of asset tokenization and registries (called rights exchanges) are not specified and deferred as implementation choices.  This scoping provides flexibility but may come at the cost of interoperability in those phases.

\textbf{Fox Verify }(\url{www.verify.fox}) is a DLT-based content licensing and provenance system, that represents assets as hierarchical NFTs (EIP-6150) and uses smart contracts for rights management. It implements the Authenticity component of the Content ARCs framework, with limited support for Rights and Compensation. For the authenticity component, content identification is achieved through  cryptographic hashing, generating unique identifiers for NFTs in conjunction with perceptual hashing via PDQ \cite{pdq2019} to perform asset lookup via a `Verify' tool to retrieve and view asset metadata.  Cryptographic signatures, linked to an on-chain identity registry of publisher key pairs, provide a verifiable attestation of content origin and ownership. Rights management is implemented through machine-readable smart contract licenses, enabling structured access and usage control. Digital assets are represented as NFTs within a `ContentGraph' smart contract, which provides a structured and extensible framework for managing rights information. The hierarchical NFT structure facilitates license inheritance from parent to child nodes, facilitating rights management across complex content collections. Fox Verify enables content monetization through smart contract-based license sales, with transactions conducted in Polygon’s MATIC token. However, it lacks passive revenue-sharing mechanisms, such as royalties for derivative works.

\textbf{Commercial startups} have emerged aligned to Content ARCs to deliver AI  contributor compensation. Bria (\url{www.bria.ai}) and ProRata (\url{www.prorata.ai}) state that they train their AI models on curated licensed content, and so the issue of Authenticity and content identification within their centralized systems is moot. For Compensation, both companies use proprietary attribution mechanisms to evaluate the relevance of training data to generated outputs. Bria distributes revenue through a scoring mechanism, compensating contributors at both training and inference time based on their data’s influence. ProRata provides credit and compensation on a per-use basis. However, neither enables granular rights management. Dataswyft (\url{www.dataswyft.com}) offers a contract-based data ecosystem facilitating self-sovereign data ownership without scalability concerns of DLT. For rights, the HAT Microserver Instruction Contracts define the terms under which external parties can access a user’s data to train AI but require manual enforcement and renewal by Dataswyft. The system facilitates compensated access to data within a contractually enforced marketplace. Other startups and image stock companies, such as Getty (\url{www.gettyimages.co.uk/ai}), have introduced their own generative AI services that claim to compensate creators whose data was used in training the models; however, due to their proprietary nature, there is a lack of public detail on the compensation model.  In general, commercial platforms focus on either or both of the rights and compensation phases of Content ARCs, rather than provenance and authenticity of data across platforms.

\section{Conclusion}
Once implemented, Content ARCs would enable end-to-end machine-readable permissions and automatic compensation throughout the AI development. A future case study could involve a content creator utilizing ARC-compliant tools to manage rights and receive compensation for their work used in AI training. This demonstrates how our framework directly addresses current policy needs, responding to the needs outlined in the UK's recent consultation on AI and copyright, which addresses transparent licensing and fair remuneration for creators in AI development \cite{UKAICopyright2025}.

No single open standard or system for content rights yet delivers fully across all phases and capabilities of the Content ARC framework.  Several technical barriers remain, of which we list the most significant.  First, the challenge of establishing registries for authenticity, ownership and licenses leads most solutions toward centralized or DLT solutions deployed on private chains due to scalability concerns. Yet this raises governance issues around liability and operation of the registry. Identity is key to representing rightsholders and licensees, yet self-sovereign identity (SSI) identity frameworks remain under-adopted, including in the creative economy. Today, most decentralized systems are leveraging the blockchain wallet address as an identifier. Ongoing legislative debate around copyright form may hinder adoption and effectiveness of rights declarations, which are reliant upon legal remedy rather than technology (\ie DRM) to ensure compliance. A further limitation is that content discovery mechanisms are largely absent today; even NFT markets were centralized search portals, indexing decentralized registries. Finally, despite a few early  scoping workshops exploring value in decentralized licensing \cite{Liddell2024,DigitalCatapult2024} there is a lack of `in the wild' feasibility studies exploring how this would map to business models based on Content ARCs---answering this will be key to creating impact in the creative economy.

\section*{Acknowledgements}

This work was supported by UKRI Grant EP/T022485/1.

\begingroup
\renewcommand{\baselinestretch}{0.90}
\small 
\bibliographystyle{IEEEtran}
\bibliography{main}

\begin{thebibliography}{10}
\providecommand{\url}[1]{#1}
\csname url@samestyle\endcsname
\providecommand{\newblock}{\relax}
\providecommand{\bibinfo}[2]{#2}
\providecommand{\BIBentrySTDinterwordspacing}{\spaceskip=0pt\relax}
\providecommand{\BIBentryALTinterwordstretchfactor}{4}
\providecommand{\BIBentryALTinterwordspacing}{\spaceskip=\fontdimen2\font plus
\BIBentryALTinterwordstretchfactor\fontdimen3\font minus \fontdimen4\font\relax}
\providecommand{\BIBforeignlanguage}[2]{{%
\expandafter\ifx\csname l@#1\endcsname\relax
\typeout{** WARNING: IEEEtran.bst: No hyphenation pattern has been}%
\typeout{** loaded for the language `#1'. Using the pattern for}%
\typeout{** the default language instead.}%
\else
\language=\csname l@#1\endcsname
\fi
#2}}
\providecommand{\BIBdecl}{\relax}
\BIBdecl

\bibitem{nyt}
\BIBentryALTinterwordspacing
M.~M. Grynbaum and R.~Mac, ``{The Times Sues OpenAI and Microsoft Over A.I. Use of Copyrighted Work},'' \emph{The New York Times}, 2023. [Online]. Available: \url{https://www.nytimes.com/2023/12/27/business/media/new-york-times-open-ai-microsoft-lawsuit.html}
\BIBentrySTDinterwordspacing

\bibitem{UKAICopyright2025}
{UK Intellectual Property Office}, ``{Consultation on Copyright and Artificial Intelligence},'' \url{https://www.gov.uk/government/consultations/copyright-and-artificial-intelligence}, February 2025.

\bibitem{ft2023aicopyright}
\BIBentryALTinterwordspacing
{Financial Times}, ``{AI copyright wars need a market solution},'' \emph{Financial Times}, 2023. [Online]. Available: \url{https://www.ft.com/content/304d660f-6cac-4e38-a6d5-d8d98f5770fb}
\BIBentrySTDinterwordspacing

\bibitem{giustina2024fair}
A.~C.~D. Giustina, ``{Fair Compensation for Copyrighted Data Used in AI Training},'' Master's Thesis, Tilburg University, 2024.

\bibitem{EUAIInvestment2025}
\BIBentryALTinterwordspacing
A.~Cranz, ``{EU} to invest €200 billion in {A}{I} development to compete with us and china,'' \emph{The Verge}, 2025. [Online]. Available: \url{https://www.theverge.com/news/609930/eu-200-billion-investment-ai-development}
\BIBentrySTDinterwordspacing

\bibitem{journallicensing}
T.~Carpenter, ``We could use a model licensing framework for scholarly content use in {A}{I} tools,'' 2025, available at: \url{https://scholarlykitchen.sspnet.org/2025/02/26/we-could-use-a-model-licensing-framework-for-ai-tools/}.

\bibitem{TDMRep2024}
{W3C}, ``{TDM Reservation Protocol (TDMRep)},'' \url{https://www.w3.org/community/reports/tdmrep/CG-FINAL-tdmrep-20240202}, February 2024.

\bibitem{EUCopyright2019}
\BIBentryALTinterwordspacing
``{Directive (EU) 2019/790 of the European Parliament and of the Council of 17 April 2019 on Copyright and Related Rights in the Digital Single Market},'' \emph{{Official Journal of the European Union}}, 2019. [Online]. Available: \url{https://eur-lex.europa.eu/legal-content/EN/TXT/?uri=CELEX%3A32019L0790}
\BIBentrySTDinterwordspacing

\bibitem{IPTC2024}
\BIBentryALTinterwordspacing
{International Press Telecommunications Council (IPTC)}, \emph{{IPTC Photo Metadata Standard v2024.1}}, 2024. [Online]. Available: \url{https://iptc.org/standards/photo-metadata/}
\BIBentrySTDinterwordspacing

\bibitem{C2PA2024}
\emph{Coalition for Content Provenance and Authenticity, Technical Specification v2.1}, \url{https://c2pa.org}, 2024.

\bibitem{trustmark}
\BIBentryALTinterwordspacing
T.~Bui, S.~Agarwal, and J.~Collomosse, ``{TrustMark: Universal Watermarking for Arbitrary Resolution Images},'' \emph{arXiv preprint arXiv:2311.18297}, 2023. [Online]. Available: \url{https://arxiv.org/abs/2311.18297}
\BIBentrySTDinterwordspacing

\bibitem{fernandez2024videoseal}
\BIBentryALTinterwordspacing
P.~Fernandez, H.~Elsahar, I.~Z. Yalniz \emph{et~al.}, ``Video seal: Open and efficient video watermarking,'' \emph{arXiv preprint arXiv:2412.09492}, 2024. [Online]. Available: \url{https://arxiv.org/abs/2412.09492}
\BIBentrySTDinterwordspacing

\bibitem{Black2021DeepImageComparator}
A.~Black, T.~Bui, H.~Jin \emph{et~al.}, ``{Deep Image Comparator: Learning to Visualize Editorial Change},'' in \emph{Proc. CVPRW (Workshop on Media Forensics)}, 2021.

\bibitem{ISCC}
``{International Standard Content Code (ISCC)},'' \url{https://www.iso.org/standard/77899.html}.

\bibitem{nguyen2021oscar}
E.~Nguyen, T.~Bui, V.~Swaminathan \emph{et~al.}, ``{OSCAR-Net: Object-centric Scene Graph Attention for Image Attribution},'' in \emph{Proc. ICCV}, 2021.

\bibitem{balan2023b}
K.~Balan, A.~Black, S.~Jenni \emph{et~al.}, ``{DECORAIT - DECentralized Opt-in/out Registry for AI Training},'' in \emph{Proc. of the Conference on Visual Media Production (CVMP)}, 2023.

\bibitem{plus}
{PLUS Coalition}, ``{PLUS Coalition},'' \url{https://www.useplus.com/}, 2020.

\bibitem{Collomosse2024}
J.~Collomosse and A.~Parsons, ``{To Authenticity, and Beyond! Building Safe and Fair Generative AI upon the Three Pillars of Provenance},'' \emph{IEEE Computer Graphics and Applications (IEEE CG\&A)}, 2024.

\bibitem{jpegtrust2025}
{P. Rixhon}, ``An update on {J}{P}{E}{G} trust,'' \url{https://cawg.io/meeting-notes/_attachments/2025-01-21/jpeg-trust-presentation.pdf}, January 2025.

\bibitem{NSA2025}
\BIBentryALTinterwordspacing
{National Security Agency}, ``Content credentials: Strengthening multimedia integrity in the generative ai era,'' Tech. Rep., Jan. 2025. [Online]. Available: \url{https://media.defense.gov/2025/Jan/29/2003634788/-1/-1/0/CSI-CONTENT-CREDENTIALS.PDF}
\BIBentrySTDinterwordspacing

\bibitem{zauner2010implementation}
\BIBentryALTinterwordspacing
C.~Zauner, ``Implementation and benchmarking of perceptual image hash functions,'' Master's thesis, Upper Austria University of Applied Sciences, Hagenberg, 2010, original pHash documentation and implementations available at \url{http://www.phash.org}. [Online]. Available: \url{https://phash.org/}
\BIBentrySTDinterwordspacing

\bibitem{ISCCwebsite}
``{International Standard Content Code (ISCC)},'' \url{https://ieps.iscc.codes/iep-0004/}.

\bibitem{photodna}
M.~Steinebach, ``An analysis of photodna,'' in \emph{Proc. of the 18th International Conference on Availability, Reliability and Security}, ser. ARES '23.\hskip 1em plus 0.5em minus 0.4em\relax New York, NY, USA: Association for Computing Machinery, 2023.

\bibitem{pdq2019}
{Facebook Open Source}, ``{PDQ:} {P}erceptual {H}ash for {I}mage {N}ear-{D}uplicate {D}etection,'' \url{https://github.com/facebook/ThreatExchange}, 2019.

\bibitem{liu2016deep}
H.~Liu, R.~Wang, S.~Shan \emph{et~al.}, ``Deep supervised hashing for fast image retrieval,'' in \emph{Proc. CVPR}, 2016.

\bibitem{CreativeCommons2008}
\BIBentryALTinterwordspacing
{Creative Commons}, ``Machine-readable metadata,'' 2008. [Online]. Available: \url{https://creativecommons.org/about/downloads/}
\BIBentrySTDinterwordspacing

\bibitem{odrl2018vocab}
\BIBentryALTinterwordspacing
R.~Iannella and V.~Rodríguez-Doncel, ``{ODRL} vocabulary \& expression 2.2,'' W3C Recommendation, February 2018. [Online]. Available: \url{https://www.w3.org/ns/odrl/2/}
\BIBentrySTDinterwordspacing

\bibitem{DigitalCatapult2024}
\BIBentryALTinterwordspacing
{Digital Catapult}, ``{DLT Field Lab Report: Emerging Futures for Tokenisation and Digital Media Rights},'' UKRI DECaDE, Tech. Rep., 2024. [Online]. Available: \url{www.digicatapult.org.uk/wp-content/uploads/2024/12/Catapult_ORAgen_Emerging_Futures_Report.pdf}
\BIBentrySTDinterwordspacing

\bibitem{Balan2023}
K.~Balan, S.~Agarwal, S.~Jenni \emph{et~al.}, ``{EKILA: Synthetic Media Provenance and Attribution for Generative Art},'' in \emph{Proc. CVPR Workshop on Media Forensics (CVPRW)}, 2023.

\bibitem{wang2023evaluating}
S.~Wang, A.~Efros, J.~Zhu \emph{et~al.}, ``Evaluating data attribution for text-to-image models,'' in \emph{Proc. ICCV}, 2023.

\bibitem{asnani2024promark}
V.~Asnani, J.~Collomosse, T.~Bui \emph{et~al.}, ``{ProMark}: Proactive diffusion watermarking for causal attribution,'' in \emph{Proc. CVPR}, 2024.

\bibitem{wang2024economic}
\BIBentryALTinterwordspacing
J.~T. Wang, Z.~Deng, H.~Chiba-Okabe \emph{et~al.}, ``An economic solution to copyright challenges of generative ai,'' \emph{arXiv preprint arXiv:2404.13964}, 2024. [Online]. Available: \url{https://arxiv.org/abs/2404.13964}
\BIBentrySTDinterwordspacing

\bibitem{Liddell2024}
F.~Liddell, E.~Tallyn, E.~Morgan \emph{et~al.}, ``{ORAgen: Exploring the Design of Attribution through Media Tokenisation},'' in \emph{Proc. ACM Designing Interactive Systems (DIS)}, 2024.

\end{thebibliography}
\endgroup

\end{multicols}

\end{document}